\documentclass[10pt,journal]{IEEEtran}

\usepackage[utf8]{inputenc}
\usepackage[T1]{fontenc}

\usepackage{graphicx}
\usepackage{physics}
\usepackage[nolist,nohyperlinks]{acronym}
\usepackage{amsmath}
\usepackage{amsthm}
\usepackage{amssymb}
\usepackage{authblk}

\usepackage{bbm}
\usepackage{subcaption}
\usepackage{cite}

\usepackage[ruled,vlined,linesnumbered]{algorithm2e}

\usepackage{tikz}
\usetikzlibrary{automata, arrows, positioning, quantikz}

\usepackage{tikz-network}
\usepackage{todonotes}
\usepackage{xcolor}

\usepackage{acronym}
\newacro{CNOT}{controlled $X$}
\newacro{GHZ}{Greenberger-Horne-Zeillinger}
\newacro{QFI}{Quantum Fisher Information}
\newacro{QFIM}{Quantum Fisher Information Matrix}

\title{On the Characterization of Quantum Flip Stars with Quantum Network Tomography}

\author[1]{Matheus Guedes de Andrade}
\author[2]{Jake Navas}
\author[2]{Inès Montaño}
\author[1]{Don Towsley}

\affil[1]{Manning College of Information and Computer Science, University of Massachusetts Amherst}
\affil[2]{Department of Applied Physics and Materials Science, Northern Arizona University}

\begin{document}

\maketitle

\begin{abstract}
    The experimental realization of quantum information systems will be difficult due to how sensitive quantum information is to noise. Overcoming this sensitivity is central to designing quantum networks capable of transmitting quantum information reliably over large distances. Moreover, the ability to characterize communication noise in quantum networks is crucial in developing network protocols capable of overcoming the effects of noise in quantum networks. In this context, quantum network tomography refers to the characterization of channel noise in a quantum network through end-to-end measurements. In this work, we propose network tomography protocols for quantum star networks formed by quantum channels characterized by a single, non-trivial Pauli operator. Our results further the end-to-end characterization of quantum bit-flip star networks by introducing tomography protocols where state distribution and measurements are designed separately. We build upon previously proposed quantum network tomography protocols, as well as provide novel methods for the unique characterization of bit-flip probabilities in stars. We introduce a theoretical benchmark based on the Quantum Fisher Information matrix to compare the efficiency of quantum network protocols. We apply our techniques to the protocols proposed, and provide an initial analysis on the potential benefits of entanglement for Quantum Network Tomography. Furthermore, we simulate the proposed protocols using NetSquid to assess the convergence properties of the estimators obtained for particular parameter regimes. Our findings show that the efficiency of protocols depend on parameter values and motivate the search for adaptive quantum network tomography protocols.
\end{abstract}
\section{Introduction}

Quantum networks are a critical component of the next quantum revolution. The interconnection of quantum processing systems with channels that provide quantum communication are key for the scalability of quantum computers~\cite{buhrman2003distributed}, and enable applications such as quantum key distribution~\cite{vlachou2018quantum, bennett2020quantum}, quantum secrete sharing~\cite{hillery1999quantum} and distributed quantum sensing~\cite{zhuang2018distributed, guo2020distributed, Zhang21}. Despite recent experimental demonstrations of entanglement distribution in quantum networks with fiber~\cite{pompili2021realization, hermans2022qubit} and free-space communications~\cite{liao2018satellite, liao2017satellite}, the fragility of quantum information in the face of noise remains as the major barrier to the physical realization of scalable, useful quantum networks. This barrier is inherent to the complexity of quantum communication systems, which must integrate diverse quantum and classical hardware. In particular, hardware imperfections introduce unavoidable noise in the quantum information exchanged among network nodes during communication. A quantum network node must be capable of initializing, storing, and processing quantum information, either using memory in the form of matter qubits~\cite{sangouard2009quantum, ruf2021quantum, dhara2022multiplexed} or by storing photons in delay lines~\cite{azuma2015all}. The inefficiencies in memory devices introduce noise in the form of decoherence and loss~\cite{hartmann2007role}, as well as through gate imperfections which can introduce diverse processing noise. Moreover, photons are the fundamental transmission media for quantum information and performing transduction is key in networks with matter-based memories that are not optically active~\cite{christensen2020coherent, PRXQuantum.2.017002}. In addition to transduction, frequency conversion is necessary since different light frequencies are optimal for different applications~\cite{PRXQuantum.2.017002}. For instance, optimal frequencies for processing can differ from the usual telecom band that reduces photon loss in fibers~\cite{rakher2010quantum}. Due to unavoidable imperfections in implementation, transduction and frequency conversion methods are themselves sources of noise. Finally, the propagation of photons in fiber and free space incurs losses and phase errors, which can corrupt the information encoded in photons~\cite{peng2005experimental, takeoka2014fundamental}.

The noise introduced in the different layers of communication hardware accumulates as multiple nodes are used for communication. Whether it be in \textit{one-way} network architectures, where quantum information is directly transmitted across quantum channels interconnecting the nodes, or in \textit{two-way} architectures, where the channels are used to generate and propagate entangled states for teleportation, the greater the number of nodes required to establish communication, the higher the noise introduced in the information transmitted. Therefore, the development of quantum error correction codes and decoders~\cite{rozpkedek2021quantum}, as well as purification protocols~\cite{dur1999quantum,riera2021entanglement} capable of improving upon the negative effects of noise in quantum communication is instrumental to further the physical implementation of useful quantum networks. Furthermore, the design of noise-aware applications is fundamental in Noisy Intermediate Scale Quantum (NISQ) hardware, since it is not possible to rely on complex quantum error correction protocols to achieve fault-tolerant quantum operations. In particular, quantum circuit compilation routines can be optimized based on gate and memory noise models to obtain a gain in performance~\cite{murali2019noise}. 

The demand for quantum error correction and purification protocols, as well as the design of noise-aware applications renders network noise characterization as a central topic in the development of quantum networks. Error decoding in quantum error correction protocols benefits vastly from the characterization of errors processes. In addition, noise-aware applications need to use either static or dynamic error models to optimize behavior and increase efficiency~\cite{murali2019noise}. In this context, \textit{Quantum Network Tomography} (QNT) has been previously introduced to address the end-to-end characterization of network links~\cite{de2022quantum}. It connects classical network tomography~\cite{HMST2021tomography} with quantum parameter estimation~\cite{helstrom1969quantum} to devise efficient characterization methods for link noise in quantum communication through end-to-end measurements. End-to-end characterization is based on the assumption that quantum network infrastructure cannot be directly accessed to estimate link parameters. Thus, network users must obtain channel estimates by measuring quantum states that were distributed through the network. End-to-end estimation is considerably harder than point-to-point link estimation, i.e., independent estimation of network links, due to the fact that quantum channels cannot be probed directly. Instead, they act as hidden, unobservable processes for which statistics must be obtained by observing the systemic behavior of the network.

QNT differs from \textit{Quantum Process Tomography} (QPT) in a meaningful way. QNT considers an end-to-end network estimation problem, where the parameters to be estimated cannot be directly measured, while QPT aims to estimate black-box processes. For simplicity, the QNT formulation focused on in this article, assumes that network links represent quantum channels, i.e \textit{Completely Positive Trace Preserving} (CPTP) maps, with a known parametric form. It is considered that each operator in the Kraus decomposition of a channel is of a given parametric form, and the goal is to estimate the parameters to characterize all the links. Such assumption is absent in the general QPT formulation which has the goal of estimating each component of transfer matrices characterizing CPTP maps. QPT can be used to characterize networks by introducing additional assumptions in its general formulation, although we do not address methods of this form in this article.


\subsection{Contributions}

Providing efficient solutions for the characterization of arbitrary quantum networks is extremely challenging. Nonetheless, the initial work in QNT~\cite{de2022quantum} studied the particular case of quantum star networks with links representing probabilistic one-qubit Pauli channels, described by a single Pauli operator and one parameter, e.g quantum bit-flip star networks with different flip probabilities in general. Two methods have been previously proposed to obtain estimators for all of the network parameters: a method that uniquely identifies the parameter vector characterizing the network with the aid of global measurements, and a method that relies on local measurements to produce two estimates for the parameter vector. In this work, we provide additional results for the characterization of star networks and improve on the analysis of the methods proposed in~\cite{de2022quantum}. Our contributions are four-folded:

\begin{itemize}
    \item We provide a new general description of QNT protocols in which state distribution and measurements are separately defined. This definition enables methods where the same state distribution protocol is used to generate different estimators based on distinct measurements performed at the end-nodes. In addition, it enables the construction of tomography protocols that combine multiple, distinct state distribution circuits and measurements to uniquely identify network parameters.
    \item We provide novel QNT protocols for the unique characterization of bit-flip stars. The protocols use both global and local quantum measurements at the network nodes to estimate parameters. Moreover, the QNT protocols proposed generalize to stars with either $Z$ or $Y$ flip channels through a change of basis in the operators used at the network nodes.
    \item We analyze the QNT protocols proposed in this article, and compare their estimation efficiencies. Our analysis is centered on the numerical evaluation of the \textit{Quantum Fisher Information Matrix} (QFIM) containing link parameters. Our findings show that estimation efficiency of the protocols proposed depend on the values of the parameters to be estimated. This dependency in parameter value is similar to the findings reported in~\cite{fujiwara2001quantum} for the estimation of single-qubit depolarizing channels. 
    \item We simulate the designed QNT protocols in four-node star networks using the discrete-event quantum network simulator NetSquid~\cite{coopmans2021netsquid}. We use simulations to numerically analyze the convergence rate of the estimators in which our QNT protocols are based on. Our findings from simulation are in accordance with the results obtained for the QFIM, and show how the different QNT protocols behave in a particular parameter regime.
\end{itemize}

The remainder of this article is organized as follows: in Section~\ref{sec:background}, we provide the necessary background knowledge to discuss our contributions; we describe the state distribution and measurement protocols used to devise QNT protocols in Section~\ref{sec:spam}; in Section~\ref{sec:tomgoraphy}, we present the tomography protocols for the unique characterization of network links; our simulation results and numerical analysis are reported in Section~\ref{sec:evaluation}; finally, we present concluding remarks in Section~\ref{sec:conclusion}.
\section{Background}\label{sec:background}

\begin{figure}
    \centering
    \includegraphics[scale=0.15]{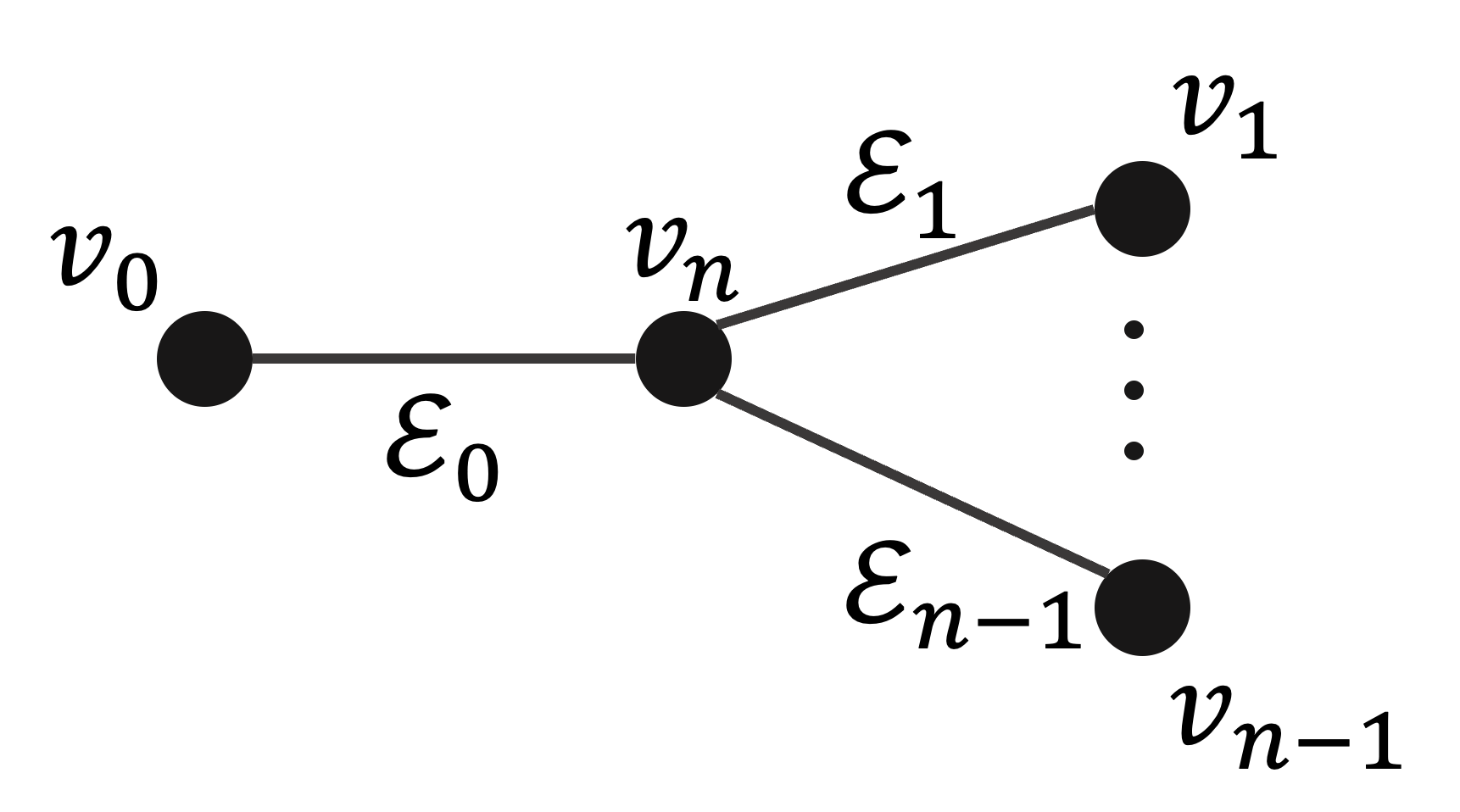}
    \caption{Quantum star networks.}
    \label{fig:star}
\end{figure}

Quantum networks are represented as graphs, with nodes representing arbitrary quantum processors and links representing quantum channels that enable the nodes to exchange quantum information. This work focuses on quantum star networks with links characterized by one-qubit probabilistic quantum channels $\mathcal{E}_e$ of the form
\begin{align}
    & \mathcal{E}_e(\rho) = \theta_e \rho + (1 - \theta_e) \sigma \rho \sigma \label{eq:singlePauli},
\end{align}
where $\theta_e \in [0, 1]$, $\rho: \mathbb{H}^{2} \to \mathbb{H}^{2}$ is the two-qubit density operator, and  $\sigma = X$. The assumption of bit-flip channels is considered for simplicity, and all the tomography protocols described in this work generalize to the case where $\sigma = Y$  or $\sigma = Z$ under a basis transformation of all the operations performed and states used. A quantum $(n + 1)$-node star network is formed by the interconnection of $n$ end-nodes through an intermediate node, as depicted in Fig.\ref{fig:star}. We represent the nodes of the star as $v_j$ for $j \in \{0, \ldots, n\}$, and label the intermediate node as $v_n$. The link $(v_j, v_n)$ represents a quantum channel $\mathcal{E}_j$ following \eqref{eq:singlePauli}.

\subsection{System model}

The nodes of the network are assumed capable of initializing qubits in the computational basis state $\ket{0}$ and of performing arbitrary quantum circuits to process them. The end-nodes communicate quantum information by preparing qubits in arbitrary states and transmitting them through the intermediate node, one qubit at a time. The channels act on the qubits transmitted through the intermediate node and corrupt the states with bit-flip noise. Furthermore, we assume that noise introduced in link $(v_j, v_n)$ by channel $\mathcal{E}_j$ is symmetric, such that the probability of a bit-flip occurring in a transmission from node $v_j$ to $v_n$ is the same as that of a flip occurring in the opposite direction. Therefore, a star network is characterized by $n$ bit-flip probabilities that specify~\eqref{eq:singlePauli} for all channels.

We consider QNT protocols for the star network using an intermediate node for quantum state distribution. Such a protocol consists of a set $\mathcal{C} = \{C_1, C_2, \ldots, C_S\}$ of state distribution circuits, $\varpi = \{\Pi_1,\ldots,\Pi_S\}$ of \textit{Positive Operator-Valued Measures} (POVM), and $\mathcal{M} = \{m_1,\ldots, m_S\}$ of number of measurement copies to be performed. In particular, the state distribution circuit $C_i$ is performed $m_i$ times to generate $m_i$ copies of a distributed state $\rho_i(\theta)$ in the end-nodes, each of which is measured with POVMs $\Pi_i$. We consider all POVMs to be projective measurements in multiple qubits. Therefore, the $M = \sum_{i}m_i$ measurement outcomes form a data set $\mathcal{D}$ of $M$ binary strings that can be used to perform estimation. It is important to emphasize that the intermediate node cannot perform quantum measurements to contribute to $\mathcal{D}$ directly, otherwise the problem reduces to the case where each channel is independently estimated. Note that circuits in $\mathcal{C}$ represent distributed circuits, and channels are used to transmit qubits when necessary.

When discussing state distribution protocols in the star, we refer to the end-node that starts the process as the \textit{root} of the protocol and to the remaining end-nodes as \textit{leaves}. Moreover, we refer to quantum state distribution circuits as quantum state distribution algorithms interchangeably in the remainder of this work.

\subsection{States in the GHZ basis}

Some of the tomography protocols proposed in this article are based on the distribution of mixed states diagonal in the Greenberger–Horne–Zeilinger (GHZ) basis. The $n$-qubit GHZ basis generalizes the Bell basis to $n$ qubits, and has $2^{n}$ states that, when written in the computational basis, assume the form
\begin{align}
    & \ket{\Phi_{s}} = \frac{\ket{0s_{1:}} + (-1)^{s_0} \ket{1\overline{s}_{1:}}}{\sqrt{2}}\label{eq:ghzInZ}
\end{align}
where $s \in \{0,1\}^{n}$ is an $n$-bit string, $\overline{s}$ is the bit-wise negation of $s$, $s_0$ is the first bit of $s$ and $s_{1:} \in \{0, 1\}^{n - 1}$ is the string obtained by removing $s_0$ from $s$. For example, when $s = 0101$, $s_0 = 0$, $s_{1:} = 101$ and $\overline{s} = 1010$. We denote the projector onto $\ket{\phi_{s}}$ as $\Phi_{s} = \dyad{\phi_s}$. Specifying the GHZ basis in the $Z$ basis is helpful when $Z$ basis measurement statistics are to be extracted from such states. Similarly, expressing the GHZ basis in the $X$ basis of $n$ qubits will prove helpful in the description of tomography protocols. As a matter of fact, a complete description of the states in the general case is not necessary and we consider the rule that specifies what states of the $X$ basis have non-zero components when $\ket{\phi_s}$ is projected in that basis. Thus, let $\ket{x^{+}}$ denote a state in the $X$ basis of $n$-qubits, such that 
\begin{equation}
    x_j^{+} =
    \begin{cases}
        +, \text{ if } x_j = 0, \\
        -, \text{ if } x_j = 1,
    \end{cases} 
\end{equation}
e.g $\ket{0^{+}1^{+}0^{+}} = \ket{+-+}$. Using the bit string representation, the inner product $\braket{x^{+}}{\phi_s}$ provides the component of $\ket{\phi_s}$ in the $X$ basis as
\begin{align}
    & \braket{x^{+}}{\phi_s} = \frac{\braket{x^{+}}{0s_{1:}} + (-1)^{s_0}\braket{x^{+}}{1\overline{s}_{1:}}}{\sqrt{2}}.
\end{align}
Through algebraic manipulations, it is possible to show that
\begin{align}
    & \braket{x^{+}}{\phi_s} = \frac{(-1)^{x \cdot 0s_{1:}} + (-1)^{s_0 + x \cdot 1\overline{s}_{1:}}}{\sqrt{2^{n + 1}}},
\end{align}
where $\cdot$ denotes the inner product between two binary strings. The inner product allows to compute the probability of measuring state $\ket{\phi_s}$ in the $X$ basis and obtaining state $\ket{x^{+}}$ as
\begin{align}
    & \abs{\braket{x^{+}}{\phi_s}}^2 = \frac{(s_0 + \tilde{1} \cdot x) \bmod{2}}{2^{n - 1}},\label{eq:xMeasureRule}
\end{align}
where $\tilde{1} = 1 \ldots 1$ denotes an $n$-bit string with all bits equal one.
This last result implies that the probability will be non-zero only when the parity of $x$ is $s_0$, i.e if the number of $-$ labels is even when $s_0 = 0$ and odd when $s_1 = 1$. The binary parity of a string $s$ will appear in the definition of different estimators for the tomography problem. Therefore, let $\beta: \{0, 1\}^{n} \to \{0, 1\}$ denote the function
\begin{align}
    \beta(s) = (\sum_{i = 0}^{n - 1} s_i)\bmod{2}.\label{eq:binparity}
\end{align}


\subsection{Quantum Parameter Estimation}\label{sec:qfim}
The quantum parameter estimation problem captures the estimation of a parameter vector $\theta \in \mathbb{R}^{n}$ from a $\theta$-dependent mixed state with density matrix $\rho(\theta)$. The goal is to describe an estimator $\hat{\theta} \in \mathbb{R}^{n}$ for $\theta$ based on measurement statistics obtained from $\rho(\theta)$.
A set of POVMs $\{\Pi_j\}$ is applied to $\rho$ to generate a data set $\mathcal{D}$ of observations that allow one to obtain statistics to estimate $\theta$. In this work, we focus on quantum parameter estimation problems with $Q$-qubit mixed states having density matrices of the form
\begin{equation}
    \rho(\theta) = \sum_{k = 0}^{2{^{Q}} - 1} \lambda_{k}(\theta) \Lambda_k,\label{eq:eigenDensity}
\end{equation}
where $\lambda_k: \mathbb{R}^{n} \to [0, 1]$ and $\Lambda_k: \mathbb{H}^{2^{Q}} \to \mathbb{H}^{2^{Q}}$ denote the $k$-th eigenvalue of $\rho(\theta)$ and the projector onto its correspondent eigenvector, respectively. Thus, the only dependence of $\rho$ in $\theta$ comes from its eigenvalues, i.e., $\rho$ depends on $\theta$ and its eigenvectors do not. Such states are the focus in this work since they arise from the action of quantum channels of the form in \eqref{eq:singlePauli}. Note that $Q$ is arbitrary in this case, although our methods are based on states with $Q = n - 1$ and $Q = n$.

The Quantum Fisher Information Matrix (QFIM) is a fundamental tool in the analysis of quantum estimation problems~\cite{liu2019quantum,helstrom1969quantum, meyer2021fisher}. For a given state $\rho(\theta)$, the entries of the QFIM have the form
\begin{align}
    & \mathcal{F}_{ij}^{\rho} = \frac{1}{2} \Tr[\rho \{L_i, L_j\}], \text{ where }\\
    & \frac{\partial \rho(\theta)}{\partial \theta_j} = \frac{1}{2} \{\rho(\theta), L_j\} \text{ for all } j,
\end{align}
and $\{A, B\}=AB + BA$ denotes the anti-commutator of operators $A$ and $B$. The matrix $L_j$ is known as the \textit{Symmetric Logarithm Derivative} (SLD) operator of parameter $\theta_j$ and its diagonal basis is the optimal basis to measure $\rho$ in order to extract statistics to estimate $\theta_j$. The element $\mathcal{F}_{jj}^{\rho}$ is known as the Quantum Fisher Information (QFI) of $\rho$ with respect to $\theta_j$, which gives how much information a measurement from $\rho$ contains about $\theta_j$. Since the SLDs specify the optimal measurements basis for each individual parameter, an optimal measurement for all parameters exists if and only if all SLDs commute with each other. Moreover, the invertibility of the QFIM specifies whether or not the entire parameter vector can be jointly estimated. In particular, estimators for $\theta$ derived from measurement statistics of $\rho$ are underdetermined if $\mathcal{F}^{\rho}$ is singular. When $\rho(\theta)$ is full-rank and has the form in \eqref{eq:eigenDensity}, the QFIM has entries
\begin{align}
    \mathcal{F}_{ij}^{\rho} = \sum_{k}\frac{1}{\lambda_k}\frac{\partial \lambda_{k}}{\partial \theta_i} \frac{\partial \lambda_{k}}{\partial \theta_j}.\label{eq:qfim_def}
\end{align}

The QFIM establishes the quantum Cramèr-Rao bound (QCRB), which is described as follows. Any estimator $\hat{\theta}$ constructed from measurement statistics of $\rho$ has a covariance matrix $\Sigma_{\hat{\theta}}: \mathbb{R}^{n} \to \mathbb{R}^{n}$ which holds the bound
\begin{equation}
    \Sigma_{\hat{\theta}} \geq (\mathcal{F}^{\rho})^{-1}.\label{eq:cramer_rao}
\end{equation}
An estimator is said to be efficient if its covariance matrix satisfies the QCRB with equality. 

\section{State distribution and measurement protocols} \label{sec:spam}

The goal of quantum network tomography is to estimate link parameters through end-to-end measurements. As any estimation process, quantum network tomography requires the encoding of parameters in quantum states, which can be measured to obtain parameter statistics. Therefore, there are three key steps that guide the analysis of quantum network tomography. It is necessary to design 1) state distribution protocols capable of generating the required parametrized states, 2) measurement protocols (POVMs) to be performed at the nodes, and 3) estimators taking measurements as inputs. Previous work defines solutions for the tomography problem where these three steps are done in unison~\cite{de2022quantum}. This work provides a more general description of network tomography protocols, considering different state distribution and measurements protocols as building blocks. In this section, we explore these building blocks and devise state distribution protocols and measurements for QNT. Without loss of generality, the analysis presented considers the case of bit-flip channels. Nonetheless, the methods are easily generalized to any other channel of the form in \eqref{eq:singlePauli}, i.e., a pure $Z$ or pure $Y$ channel, through a change of basis.

\subsection{States and measurement probabilities in parametric forms}

It is convenient to describe states in parametric forms for the analysis presented in this section. Let $\rho(\theta)$ be an $n$-qubit density matrix that depends on parameter $\theta$. The states of interest are of the parametric form shown in \eqref{eq:eigenDensity},
where the eigenvalues of the density matrix depend on $\theta$ and the eigenvectors do not. Such states can be represented by describing a probability function $p_\rho: \{0, 1\}^{n} \to [0, 1]$ that maps the binary label of an eigenvector to its corresponding eigenvalue. In particular, $p_\rho(s, \theta) = \lambda_k(\theta)$, where $s$ is the binary representation of the integer $k$. More precisely, $\Lambda_k$ is itself a label representing a vector in a given basis $\Lambda$ over the Hilbert space of $n$ qubits. Thus, every index $k \in \mathbb{Z}^{+}$ can be understood as an $n$-bit string $s \in \{0, 1\}^{n}$ uniquely specifying a vector in the basis. Once basis $\Lambda$ is specified together with an ordering for its states, a density matrix diagonal in $\Lambda$ can be represented by the parametric function $p_{\rho}(s, \theta)$. This characterization is useful for describing the probability distribution of projective measurement outcomes of any state in an arbitrary basis. Applying the Born rule, the probability distribution of measurement outcomes of a state $\rho$ in basis $B=\{\dyad{b_i}\}$ has the form
\begin{align}
    & p^{B}_{\rho}(s, \theta) = \sum_{s' \in \{0,1\}^{n}} p(s', \theta) \abs{\braket{b_i}{\Lambda_k}}^2\label{eq:born},
\end{align}
where $s$ and $s'$ are the binary representations of integers $i$ and $k$, respectively.

Parametric forms of state eigenvalues yield the description of parametric forms for their measurement probabilities in arbitrary bases.
This description is interesting as it provides a path for parameter estimation based on measurements of a state in different bases. Throughout the remainder of this article, we use the notation $p_{\mathcal{C}}^{B}(s, \theta)$ to denote the probability of measuring label $s \in \{0, 1\}^{n}$ from a $B$-basis projective measurement performed in a state $\rho(\theta)$ distributed by a quantum circuit $\mathcal{C}$. Moreover, $p_{\mathcal{C}}(s, \theta)$ denotes measurement probabilities in the eigenbasis of $\rho(\theta)$ distributed by $\mathcal{C}$. It is helpful to define the probability distribution function $\alpha: \{0,1\}^{n} \times [0,1]^{n} \to [0,1]$ in the form
\begin{equation}
   \alpha(s, \theta) = \prod_{j = 0}^{n - 1} \overline{s}_j\theta_j +  s_j(1 - \theta_j)\label{eq:jointIndDist},
\end{equation}
which represents a joint probability distribution of $n$ independent binary random variables. In particular, $\theta_j$ and $(1 - \theta_j)$ are the probabilities of observing the $j$-th bit of $s$ as $s_j = 0$ and $s_j = 1$, respectively. The form of $\alpha$ provides simple estimators for $\theta$. In particular, let $S \in \{0, 1\}^{n}$ denote an $n$-bit random variable, with $S_j \in \{0, 1\}$ denoting the its $j$-th random bit. It is possible to write an estimator for $\theta_j$ in the form $\hat{\theta}_j = \hat{Pr}[S_j = 0]$,
where $\hat{Pr}[S_j = 0]$ is any estimator for the probability that $S_j = 0$.

\subsection{Encoding network parameters in quantum states}

Remote state distribution is used in this work to generate quantum states of interest. Distribution of a quantum state from the root to the leaves can be described in terms of a distributed circuit implemented by the nodes of the network. The nodes initialize quantum registers at the beginning of the distribution process. The links are used to communicate quantum information, which manifests either as the direct transmission of qubits (one-way architecture), or as the generation of Bell pairs between two nodes (two-way architecture). Local quantum operations are performed at the nodes, progressively transforming the joint initialized state into the desired output. By using network links for communication, the final distributed states depend on channel parameters and allow for parameter inference through measurements.

\begin{figure*}
    \begin{centering}
        \subfloat[Network.\label{subfig:network}]{\includegraphics[scale=0.12]{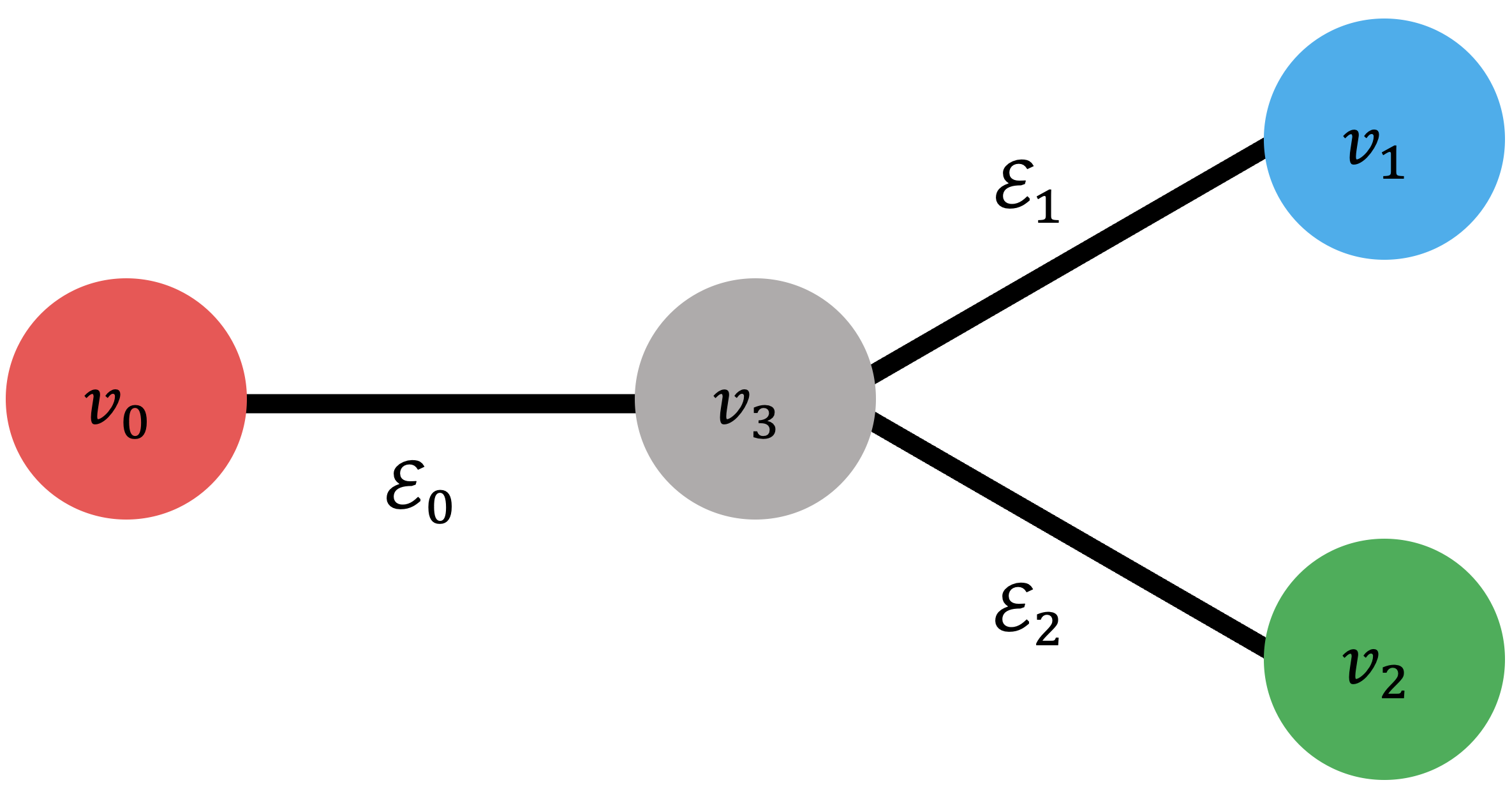}} \hfill
        \subfloat[Multicast.\label{subfig:multcirc}] {\includegraphics[scale=0.35]{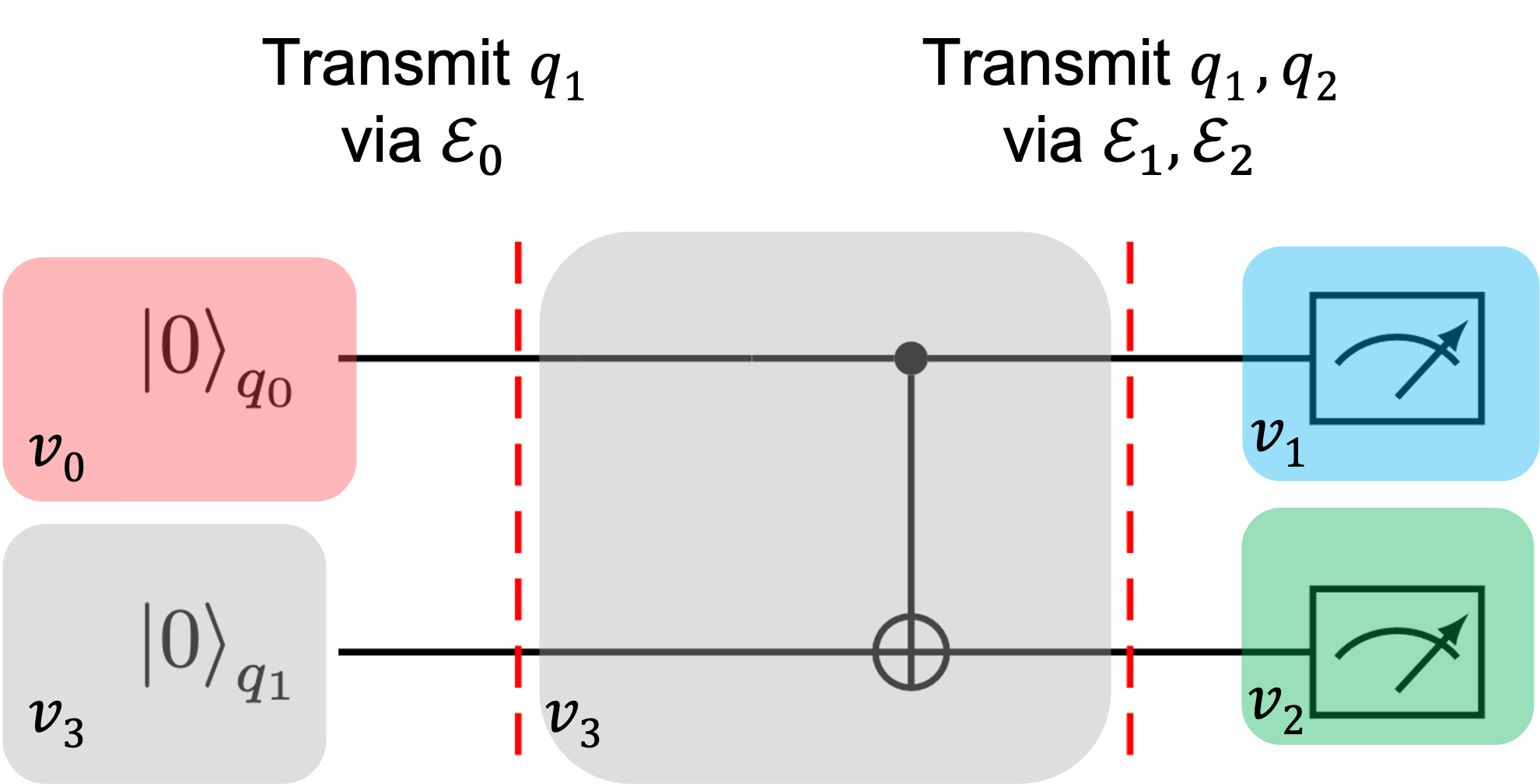}} \hfill
        \subfloat[Root-independent. \label{subfig:ricirc}]{\includegraphics[scale=0.35]{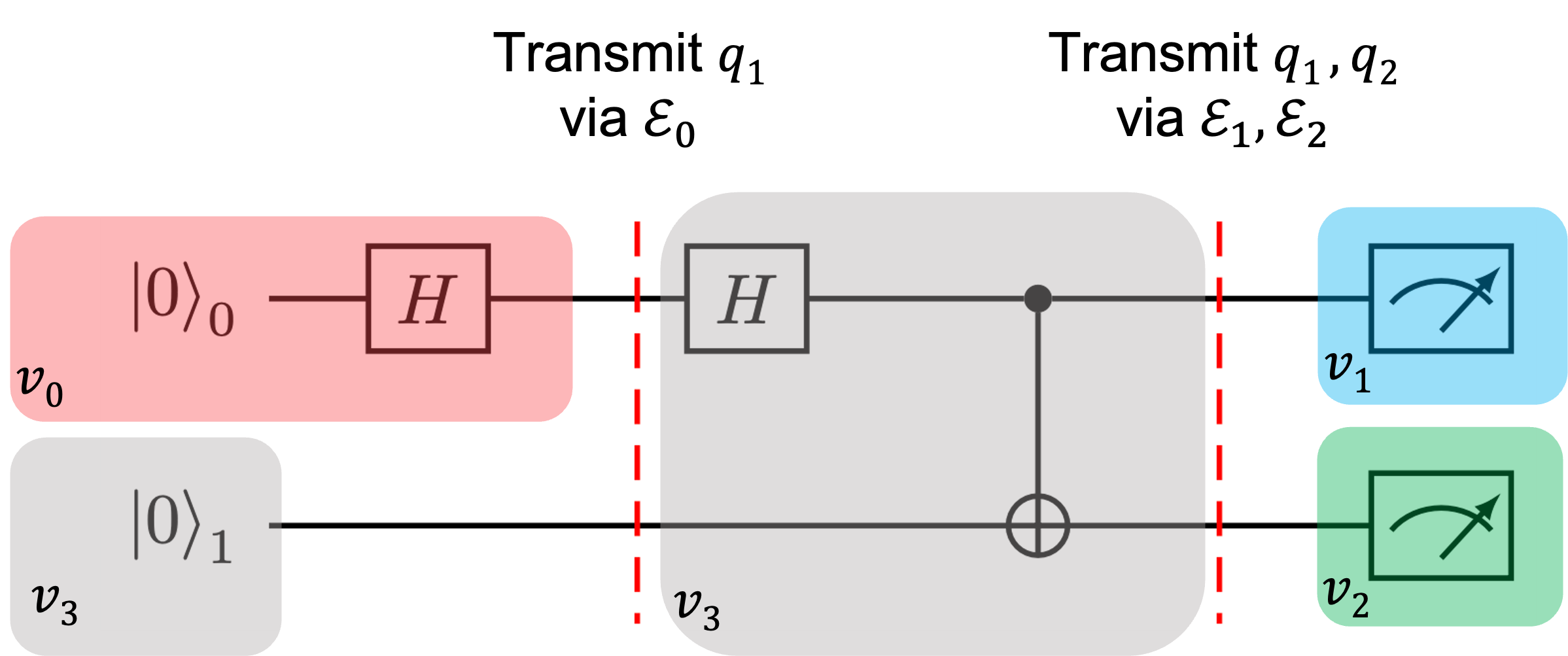}} \hfill \\
        \hfill
        \subfloat[Independent Encoding.\label{subfig:iecirc}]{\includegraphics[scale=0.35]{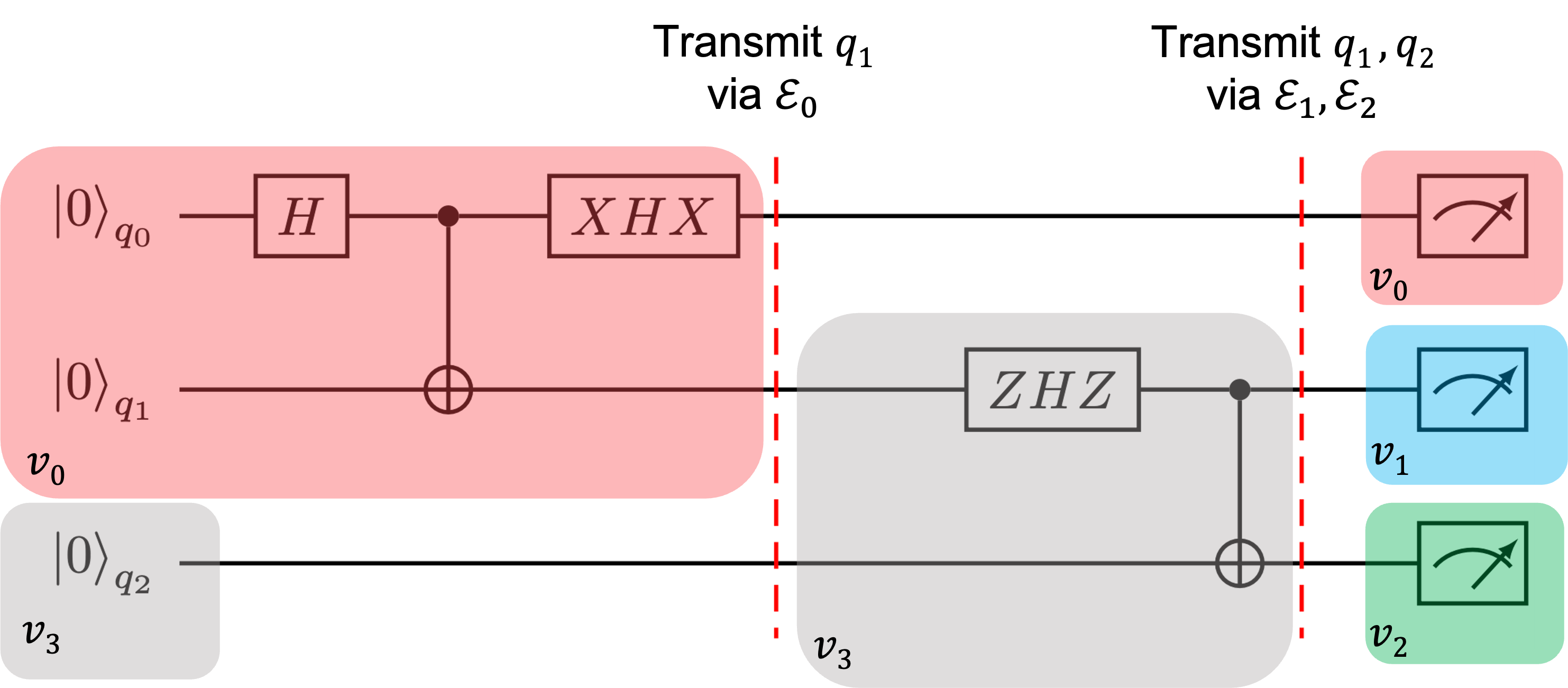}} \hfill
        \subfloat[Back-and-forth. \label{subfig:bfcirc}]{\includegraphics[scale=0.35]{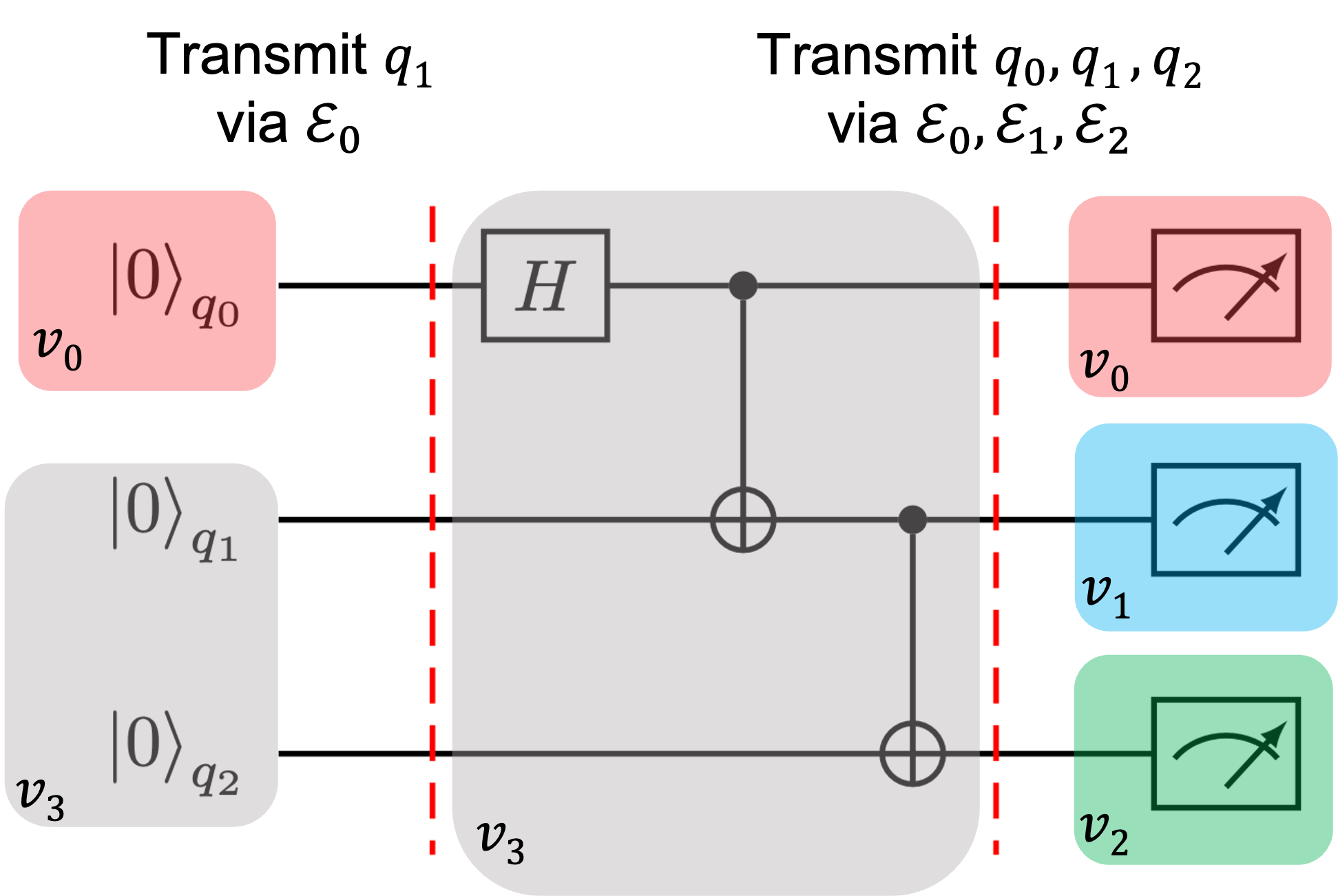}} \hfill
    \end{centering}
    \caption{State Distribution Circuits. Each circuit represents the operations performed at the nodes of a four-node star (\subref{subfig:network}) to distribute states for tomography. Colored blocks represent different nodes in the star. The gray block denotes node $v_n$, while the red, blue and green blocks denote nodes $v_0, v_1,$ and $v_2$, respectively. Vertical red-dashed lines denote the transmission of qubits through the channels in the network.}
    \label{fig:circuits}
\end{figure*}


\subsubsection{Previous state distribution protocols}

In previous work, a general state distribution algorithm for network tomography was defined under the restriction of a single channel use for each distributed state~\cite{de2022quantum}. The algorithm generated two distribution circuits for the solution of tomography problems in star networks, which are of interest to this  work. For the sake of completeness, we explicitly present the two distribution circuits defined in~\cite{de2022quantum}. For both circuits, consider $v_0$ to be the node selected to initiate state distribution, i.e., the root.

The first distribution circuit, which is referred to as the \textit{Multicast} (M) circuit in this work, distributes $Z$-diagonal states. Node $v_0$ prepares a qubit in the pure-state $\ket{0}$ and sends it to node $v_n$. Node $v_n$ receives a qubit in a mixed state given the action of channel $\mathcal{E}_0$. Then, a multi-target CNOT gate is performed in $v_n$, using the received mixed state as control and $n - 2$ newly initialized qubits in state $\ket{0}$ as targets. Each of the outputs is transmitted to a leaf of the star. The eigenvalues of the $(n-1)$-qubit density matrix describing the state of the qubits in the end-nodes have the parametric form
\begin{align}
    & p_{M}(s, \theta) = \theta_0 \alpha(s, \theta_{1:}) + (1 - \theta_0) \alpha(\overline{s}, \theta_{1:}), \label{eq:multiProb}
\end{align}
where $s \in \{0,1\}^{n - 1}$, $\theta_{1:} \in \mathbb{R}^{n - 1}$ is the vector $[\theta_1, \ldots, \theta_{n - 1}]$, and $\alpha$ is given in \eqref{eq:jointIndDist}. We expand $p_{M}$ for a four-node star in Table \ref{tab:zdiag} and show the Multicast distribution circuit for such a star in Fig.\ref{subfig:multcirc}.

\begin{table}
    \centering
    \begin{tabular}{|c|c|c|c|}
        \hline &&\\[-1em]
        State & Label & $p_{M}$ & $p_{RI}$ \\
        \hline &&\\[-1em]
        $\ket{00}$ & 00 & $\theta_0 \theta_1 \theta_2 + \overline{\theta}_0\overline{\theta}_1\overline{\theta}_2$ & $\theta_1 \theta_2$ \\ 
        \hline &&\\[-1em]
        $\ket{01}$ & 01 & $\theta_0 \theta_1 \overline{\theta}_2 + \overline{\theta}_0\overline{\theta}_1\theta_2$ & $\theta_1 \overline{\theta}_2$ \\
        \hline &&\\[-1em]
        $\ket{10}$ & 10 & $\theta_0 \overline{\theta}_1 \theta_2 + \overline{\theta}_0\theta_1\overline{\theta}_2$ & $\overline{\theta}_1 \theta_2$ \\
       \hline &&\\[-1em]
       $\ket{11}$ & 11 & $\theta_0 \overline{\theta}_1\overline{\theta}_2 + \overline{\theta}_0\theta_1\theta_2$ & $\overline{\theta}_1 \overline{\theta}_2$\\
        \hline 
    \end{tabular}
    \caption{Probability distribution of eigenbasis measurements for $Z$-diagonal states. For clarity, we denote $(1 - \theta_j)$ as $\overline{\theta}_j$.}
    \label{tab:zdiag}
\end{table}

The second algorithm distributes GHZ-diagonal states and is similar to the first with modifications in the operations performed in $v_0$ and $v_n$. We refer to this circuit as the \textit{Independent Encoding} (IE) circuit. Node $v_0$ starts the process by creating two qubits in the pure Bell state $\ket{\Phi_{00}}$. The one-qubit gate $XHX$ is applied to one of the qubits, while the other is sent to $v_n$ through channel $\mathcal{E}_0$. The qubit received in $v_n$ is operated on with the one-qubit gate $ZHZ$, followed by a mutli-target CNOT gate similar to the one used for the M algorithm. The outputs of these gates are sent to the leaves of the network. The GHZ-diagonal $n$-qubit state in the leaves of the network has eigenvalues of the form
\begin{align}
    & p_{IE}(s,\theta) = \alpha(s, \theta), \label{eq:ghzIndependentProb}
\end{align}
which are fully expanded in Table \ref{tab:ghzdiag} for a four-node star. The distribution circuit for this particular case is depicted in Fig.\ref{subfig:iecirc}. Note that the probability function in \eqref{eq:ghzIndependentProb} has the form of a joint probability distribution of independent binary random variables, and the probabilities shown in the third column of Table~\ref{tab:ghzdiag} can be interpreted as follows: the label of each state is a three-qubit measurement outcome; for the $i$-th qubit, the probabilities that the measured bit is one and zero are $\theta_i$ and $(1 - \theta_i)$, respectively; finally, the joint measurement probability is the product of the individual probabilities, which comes from a global GHZ measurement at the end-nodes. 

The examples shown in Tables \ref{tab:zdiag} and \ref{tab:ghzdiag} demonstrate the differences in parameter encoding obtained by distinct distribution circuits. Such differences manifest themselves in the findings reported in \cite{de2022quantum}, where estimators based on the M circuit were not able to uniquely determine parameters, while the ones based on the IE circuit were. These findings are revisited in Section~\ref{sec:tomgoraphy}, where novel estimators combining different states and measurements are introduced.

\begin{table}
    \centering
    \begin{tabular}{|c|c|c|c|}
        \hline &&\\[-1em]
        State & Label & $p_{IE}$ & $p_{BF}$ \\
        \hline &&\\[-1em]
        $\ket{000} + \ket{111}$ & 000 & $\theta_0 \theta_1 \theta_2$& $\theta_0\left(\theta_0\theta_1\theta_2\
        +\Bar{\theta_0}\Bar{\theta_1}\Bar{\theta_2}\right)$ \\
        \hline &&\\[-1em]
        $\ket{001} + \ket{110}$ & 001 & $\theta_0 \theta_1 \overline{\theta}_2$&
        $\theta_0\left(\theta_0\theta_1\bar{\theta_2} +\Bar{\theta_0}\Bar{\theta_1}\theta_2\right)$ \\
        \hline &&\\[-1em]
        $\ket{010} + \ket{101}$ & 010 & $\theta_0 \overline{\theta}_1 \theta_2$&
        $\theta_0\left(\theta_0\bar{\theta_1}\theta_2 +\Bar{\theta_0}\theta_1\Bar{\theta_2}\right)$\\
        \hline &&\\[-1em]
        $\ket{011} + \ket{100}$ & 011 & $\theta_0 \overline{\theta}_1 \overline{\theta}_2$&
        $\theta_0\left(\theta_0\bar{\theta_1}\bar{\theta_2} +\Bar{\theta_0}\theta_1\theta_2\right)$\\
        \hline &&\\[-1em]
        $\ket{000} - \ket{111}$ & 100 & $\overline{\theta}_0 \theta_1 \theta_2$&
        $\Bar{\theta_0}\left(\theta_0\theta_1\theta_2 +\Bar{\theta_0}\Bar{\theta_1}\Bar{\theta_2}\right)$\\
        \hline &&\\[-1em]
        $\ket{001} - \ket{110}$ & 101 & $\overline{\theta}_0 \theta_1 \overline{\theta}_2$&
        $\Bar{\theta_0}\left(\theta_0\theta_1\Bar{\theta_2} +\Bar{\theta_0}\Bar{\theta_1}\theta_2\right)$\\
        \hline &&\\[-1em]
        $\ket{010} - \ket{101}$ & 110 & $\overline{\theta}_0 \overline{\theta}_1 \theta_2$&
        $\Bar{\theta_0}\left(\bar{\theta_0\theta_1}\theta_2 +\Bar{\theta_0}\theta_1\Bar{\theta_2}\right)$\\
        \hline &&\\[-1em]
        $\ket{011} - \ket{100}$ & 111 & $\overline{\theta}_0 \overline{\theta}_1 \overline{\theta}_2$&
        $\Bar{\theta_0}\left(\theta_0\bar{\theta_1}\bar{\theta_2} +\Bar{\theta_0}\theta_1\theta_2\right)$\\
        \hline 
    \end{tabular}
    \caption{Probability distribution of eigenbasis measurements for GHZ-diagonal states. The $\sqrt{2}$ normalization factor of the states in the GHZ basis is omitted for clarity. }
    \label{tab:ghzdiag}
\end{table}

\subsubsection{New states for parameter encoding}
We now present two new state distribution algorithms for tomography. The first algorithm is denoted as the \textit{Root Independent} (RI) algorithm, which is based on the general distribution protocol defined in~\cite{de2022quantum} when applied to star networks. The root node starts by initializing a qubit in the pure state $\ket{+}$ and transmits it to the intermediate node. The channel connecting the root to the intermediate node does not change the state of this qubit, since $\ket{+}$ is an eigenvector of $X$. Therefore, the intermediate node receives the pure state $\ket{+}$, which undergoes the action of a Hadamard gate and a generalized $(n - 1)$-qubit CNOT gate. Once the operations are finished, each qubit is sent to a leaf of the star. The RI distribution algorithm yields a $Z$-basis diagonal, $(n-1)$-qubit state of the form
\begin{align}
    & p_{RI}(s, \theta) = \alpha(s_{1:},\theta_{1:}),\label{eq:rootIndependentProb}
\end{align}
which is similar to \eqref{eq:ghzIndependentProb}, \textit{i.e} joint distribution of independent variables, although with no dependency on $\theta_0$. This property motivates the name of the algorithm, since $\theta_0$ is the parameter defining channel $\mathcal{E}_0$ that interconnects the root to the intermediate node. The probability distribution in \eqref{eq:rootIndependentProb} is exemplified in the fourth column of Table~\ref{tab:zdiag} for a four-node star.



The second algorithm is denoted as the \textit{Back-and-Forth} (BF) distribution circuit. In this protocol, the root transmits a qubit to node $v_n$ initialized in the pure-state $\ket{0}$. The intermediate node performs a GHZ generation circuit, applying a Hadarmard gate to the qubit it received and using it as the control of a multi-target CNOT gate. The output of the circuit is the GHZ-diagonal state
\begin{align}
    & \rho(\theta_0) = \theta_{0} \Phi_{00\ldots0} + (1 - \theta_0) \Phi_{10\ldots0}.
\end{align}
If a bit-flip occurs on the initial channel, the qubit received in the intermediate node is in state $\ket{1}$. After the Hadarmard gate, the control used in the CNOT gate is in state $\ket{-}$ and the output GHZ state will have a negative relative phase, i.e the state $\Phi_{10\ldots0}$. When a bit-flip does not occur, there is no relative phase in the output GHZ state, i.e the state $\Phi_{00\ldots0}$. The qubit used for control is then sent back through the initial channel, and each remaining qubit is sent to a particular leaf of the network through its respective link. The eigenvalues of the GHZ-diagonal state in the end-nodes of the network have the form
\begin{align}
    & p_{BF}(s, \theta) = \alpha(s_0, \theta_0) p_M(s, \theta),\label{eq:bfprob}
\end{align}
where $p_M(s, \theta)$ is given in \eqref{eq:jointIndDist}. The BF distribution circuit for a four-node star is shown in Fig.\ref{subfig:bfcirc}, and \eqref{eq:bfprob} is fully expanded in the fourth column of Table~\ref{tab:ghzdiag} for this case.

\subsection{Measurement protocols}

State distribution protocols are the first ingredient of quantum tomography protocols, since they provide quantum states that depend on channel parameters. In order to assess the information contained in such states, it is necessary to perform quantum measurements. In the above cases, these measurements refer to POVMs performed in the end nodes of the star. Following the discussion presented in Section~\ref{sec:qfim}, the optimal POVMs to extract statistics for estimation are projective measurements in the diagonal bases of the distributed states, which are the bases that diagonalize the corresponding SLD opertors. States $\rho_{IE}$ and $\rho_{BF}$ are diagonal in the GHZ basis, and the corresponding optimal measurements are global GHZ-basis projective measurements in the end-nodes. Such measurements are significantly more challenging than local ones, as they require distributed entanglement to be performed. Therefore, it is of interest to consider alternative non-optimal local measurements to obtain statistics for estimation when such states are considered. We now describe 1) optimal measurements for the states presented and 2) local measurement strategies for $\rho_{IE}$ and $\rho_{BF}$.


\subsubsection{Optimal measurements} The optimal measurement bases for the states are shown in Table~\ref{tab:sldMeasurements}. Since the states distributed by the circuits described in the previous section were expressed in their respective diagonal bases, the probability distribution for optimal measurements were already specified by the functions $p_{M},p_{IE}, p_{RI},$ and $p_{BF}$.

\begin{table}
    \centering
    \begin{tabular}{|c|c|c|}
        \hline
        \textbf{\textit{State distribution protocol}} & \textbf{\textit{Basis}} & \textbf{\textit{Locality}} \\
        \hline Multicast (M) \cite{de2022quantum} & $Z$ basis & Local \\
        \hline Independent encoding (IE) \cite{de2022quantum} & GHZ basis & Global \\
        \hline Root-independent encoding (RI) & $Z$ basis & Local \\
        \hline Back-and-forth (BF) & GHZ basis & Global \\
        \hline 
    \end{tabular}
    \caption{Optimal measurement basis for state distribution protocols.}
    \label{tab:sldMeasurements}
\end{table}


\subsubsection{Local measurements} The first strategy for both IE and BF is to measure the qubits in the end-nodes of the star in the $Z$ basis. Computing the measurement probability distributions for the states when all qubits are locally measured in the $Z$ basis is straightforward, since the parametric form of their eigenvalues was expressed by writing the GHZ basis in the $Z$ basis following~\eqref{eq:ghzInZ}. The measurement probability distribution for the state distributed by the IE protocol is obtained from \eqref{eq:ghzIndependentProb} and has the form
\begin{align}
    & p_{IE}^{Z}(s,\theta) = \frac{1}{2} p_{RI}(s, \theta), \label{eq:ghzIndependentProbZ}
\end{align}
with $p_{RI}$ given in~\eqref{eq:rootIndependentProb}. This measurement distribution implies that statistics obtained from local $Z$ measurements in $\rho_{IE}(\theta)$ do not depend on $\theta_0$.
\begin{table}
    \centering
    \begin{tabular}{|c|c|c|c|}
        \hline &&\\[-1em]
        State & Label & $p_{IE}^{Z}$ & $p_{BF}^{Z}$\\
        \hline &&\\[-1em]
        $\ket{000}$ & 000 & $ \theta_1 \theta_2 / 2$&
        $\left(\theta_0\theta_1\theta_2 
        +\Bar{\theta_0}\Bar{\theta_1}\Bar{\theta_2}\right) / 2$\\
        \hline &&\\[-1em]
        $\ket{001}$ & 001 & $ \theta_1 \overline{\theta}_2 / 2$&
        $\left(\theta_0\theta_1\bar{\theta_2} +\Bar{\theta_0}\Bar{\theta_1}\theta_2\right) / 2$\\
        \hline &&\\[-1em]
        $\ket{010}$ & 010 & $ \overline{\theta}_1 \theta_2 / 2$&
        $\left(\theta_0\bar{\theta_1}\theta_2 +\Bar{\theta_0}\theta_1\Bar{\theta_2}\right) / 2$\\
        \hline &&\\[-1em]
        $\ket{011}$ & 011 & $ \overline{\theta}_1 \overline{\theta}_2 / 2$&
        $\left(\theta_0\bar{\theta_1}\bar{\theta_2} +\Bar{\theta_0}\theta_1\theta_2\right) / 2$\\
        \hline &&\\[-1em]
        $\ket{100}$ & 100 & $ \theta_1 \theta_2 / 2$&
        $\left(\theta_0\bar{\theta_1}\bar{\theta_2} +\Bar{\theta_0}\theta_1\theta_2\right) / 2$\\
        \hline &&\\[-1em]
        $\ket{101}$ & 101 & $ \theta_1 \overline{\theta}_2 / 2$&
        $\left(\theta_0\bar{\theta_1}\theta_2 +\Bar{\theta_0}\theta_1\Bar{\theta_2}\right) / 2$\\
        \hline &&\\[-1em]
        $\ket{110}$ & 110 & $ \overline{\theta}_1 \theta_2 / 2$&
        $\left(\theta_0\theta_1\bar{\theta_2} +\Bar{\theta_0}\Bar{\theta_1}\theta_2\right) / 2$\\
        \hline &&\\[-1em] 
        $\ket{111}$ & 111 & $ \overline{\theta}_1 \overline{\theta}_2 / 2$&
        $\left(\theta_0\theta_1\theta_2 +\Bar{\theta_0}\Bar{\theta_1}\Bar{\theta_2}\right) / 2$\\
        \hline 
    \end{tabular}
    \caption{Probability distribution for $Z$-basis measurements for GHZ-diagonal states in a four-node star.}
    \label{tab:ghzdiagZ}
\end{table}
Similarly, the $Z$-basis measurement probability for the BF state is derived from~\eqref{eq:bfprob} and has the form
\begin{align}
    & p_{BF}^{Z}(s, \theta) = \frac{1}{2}p_{M}(s,\theta),\label{eq:bfmeasurementZ}
\end{align}
where $p_{M}$ is given in \eqref{eq:jointIndDist}. It is of interest to point out that the dependency of $p_{BF}^{Z}$ in $\theta$ is qualitatively the same as that of $p_{M}$. Hence, any estimator based on $p_{M}$, such as the one specified in~\cite{de2022quantum}, can use estimates for $p_{BF}^{Z}$ to estimate $\theta$.

The second strategy is to locally measure the qubits at the end-nodes in the $X$ basis. The probabilities in \eqref{eq:xMeasureRule} can be used with~\eqref{eq:born} to describe the measurement probabilities for both IE and BF states. Interestingly, the two states have equal measurement probabilities in the $X$ basis. Let $s \in \{0, 1\}$ denote the label of the measurement outcome in the $X$ basis. The measurement probability function has the form
\begin{align}
    & p_{IE}^{X}(s, \theta) = p_{BF}^{X}(s, \theta) = \frac{1}{4} \alpha(\beta(s), \theta_0),\label{eq:ghzProbX}
\end{align}
which depends exclusively on $\theta_0$, where $\beta(s)$ is the parity function given in \eqref{eq:binparity}. This dependency shows that $\theta_0$ can be directly estimated from $X$-basis measurement outcomes of either state and, thus, that the first channel can be characterized from such measurements.

\begin{table}
    \centering
    \begin{tabular}{|c|c|c|}
        \hline &&\\[-1em]
        State & Label & $p_{IE}^{X} = p_{BF}^{X}$ \\
        \hline &&\\[-1em]
        $\ket{+++}$ & 000 & $ \theta_0 / 4$ \\
        \hline &&\\[-1em]
        $\ket{++-}$ & 001 & $ \overline{\theta}_0 / 4$ \\
        \hline &&\\[-1em]
        $\ket{+-+}$ & 010 & $ \overline{\theta}_0 / 4$ \\
        \hline &&\\[-1em]
        $\ket{+--}$ & 011 & $ \theta_0 / 4$ \\
        \hline &&\\[-1em]
        $\ket{-++}$ & 100 & $ \overline{\theta}_0 / 4$ \\
        \hline &&\\[-1em]
        $\ket{-+-}$ & 101 & $ \theta_0 / 4$ \\
        \hline &&\\[-1em]
        $\ket{--+}$ & 110 & $ \theta_0 / 4$ \\
        \hline &&\\[-1em]
        $\ket{---}$ & 111 & $ \overline{\theta}_0 / 4$ \\
        \hline 
    \end{tabular}
    \caption{Probability distribution for $X$-basis measurements for the GHZ-diagonal states in a four-node star.}
    \label{tab:ghzdiagX}
\end{table}

\section{Quantum Network Tomography Protocols} \label{sec:tomgoraphy} 

The state distribution and measurement protocols defined in Section~\ref{sec:spam} are the necessary ingredients to characterize bit-flip probabilities in star networks. The protocols are combined in this section to construct complete tomography protocols. We now present multiple protocols that use different combinations of state distribution and measurements to provide estimators for all $n$-channel parameters in an arbitrary $(n+1)$-node star with different efficiencies. Moreover, the estimators presented are evaluated in Section~\ref{sec:evaluation} both analytically, through their respective QFIMs, and numerically, with the aid of simulation.

We categorize the tomography protocols presented in this section based on the number of distinct state distribution protocols and measurements used to obtain a unique description of the entire parameter vector. In particular, an estimator that requires a number $S$ of state distribution circuits and $P$ of measurement protocols is an $(S \cdot P)$-step protocol.

\subsection{Quantum network tomography and parameter estimation}
In order to discuss the quantum network tomography protocols proposed in this section, it is instrumental to formally address how they are framed in the theory of quantum estimation. Complete network tomography protocols require, in the general case, multiple copies of distinct states to be distributed and measured with various protocols. In quantum estimation problems, a parameter-dependent state $\rho(\theta)$ is measured in order to estimate $\theta$. We now discuss how the notion of multiple estimation steps fits into this perspective. Throughout this discussion, we use $\rho^{*}$ to denote the joint state of multiple copies of possibly different, multi-qubit quantum states that are the input to the quantum estimation problem. The dependency of $\rho^{*}$ with $\theta$ is omitted for simplicity.

We start with the analysis of one-step protocols, described by a single $Q$-qubit distribution circuit that generates a state $\rho(\theta)$, and a single measurement protocol executed at the end-nodes. Let $m$ be the number of copies of $\rho(\theta)$ to be distributed to the end-nodes to obtain parameter-dependent statistics. The goal is to construct an estimator $\hat{\theta} \in \mathbb{R}^{n}$ for $\theta \in \mathbb{R}^{n}$ using a state $\rho^{*}$ of the form
\begin{align}
    \rho^{*} = \rho(\theta)^{\otimes m},
\end{align}
where the superscript $\otimes m$ denotes the tensor product of $\rho(\theta)$ with itself $m$ times. Note that each copy of $\rho(\theta)$ is a $Q$-qubit state spread across end-nodes. Consider that the projective measurement chosen for this one-step tomography protocol is determined by projectors $\{\Pi_1, \ldots, \Pi_{2^{Q}}\}$, with
\begin{align}
    & \Pi_i = \dyad{\pi_i}.
\end{align}
The dataset of observations to obtain statistics for the construction of $\hat{\theta}$ has the form $\{s_{1},\ldots, s_{m}\}$, where each $s_{i} \in \{0, 1\}^{Q}$ is the $Q$-bit classical label denoting the $i$-th measurement outcome. It follows directly from the properties of the QFIM of separable states~\cite{liu2019quantum} that the QFIM of $\rho^{*}$ has the form
\begin{equation}
    \mathcal{F}_{\rho*} = m \mathcal{F}_{\rho(\theta)}.
\end{equation}
Note that the measurement protocols described in Section~\ref{sec:spam} measure each copy of $\rho(\theta)$ independently. Therefore, no entanglement is used across distinct copies of $\rho(\theta)$. Understanding the additional power provided by the use of entanglement across different copies to obtain estimators is a promising research direction for future work. 

The one-step case is crucial as it serves as the basis for the discussion of the general $S \cdot P$-step estimation case introduced earlier. In particular, the combination of a distributed state and a measurement protocol produces a state. Such a state is directly obtained by applying a CPTP map that corresponds to a measurement operation of the distributed state in the specified basis. Hence, consider a tomography protocol that uses one circuit $\mathcal{C}$ to distribute a $Q$-qubit state and two measurement strategies $\varpi_1 = \{\Pi_1^{1},\ldots, \Pi_{2^Q}^1\}$ and $\varpi_2 = \{\Pi_1^{2},\ldots, \Pi_{2^Q}^{2}\}$ to uniquely determine the parameters. We can represent the two scenarios by considering states of the form
\begin{equation}
    \rho_{j}(\theta) = \sum_{q = 0}^{2^{Q} - 1} \Pi_q^{j} \rho(\theta) \Pi_q^{j} = \sum_{q = 0}^{2^{Q} - 1}  \bra{\pi_{q}^{j}}\rho(\theta)\ket{\pi_{q}^{j}} \Pi_q^{j} \label{eq:diagAfterMeasurement}
\end{equation}
$j \in \{1,2\}$, and for which the QFIM can be computed. 
In this case, the QFIM is the Classical Fisher Information Matrix (CFIM) obtained using the probability distribution of measurements in the $\varpi_j$ basis.
Since projective measurements are considered, the QFIM is directly obtained from the scalar values $\bra{\pi_{q}^{j}}\rho(\theta)\ket{\pi_{q}^{j}}$, which are the probabilities in \eqref{eq:born}. Mappings following \eqref{eq:diagAfterMeasurement} are helpful when multiple copies are considered. Suppose that $m_1$ and $m_2$ copies of $\rho(\theta)$ are measured in the $\varpi_1$ and $\varpi_2$ bases, respectively. The state $\rho^{*}$ that characterizes the combined distributed copies has the form
\begin{equation}
    \rho^{*} = \rho_1(\theta)^{\otimes m_1} \bigotimes \rho_2(\theta)^{\otimes m_2},
\end{equation}
and the dataset $\mathcal{D} = \{\{s_{1}^{1},\ldots, s_{m_1}^{1}\}, \{s_1^{2},\ldots, s_{m_2}^{2}\}\}$ of measurement observations has a total of $m_1 + m_2$ entries with a sub-component for each state. In this case, the QFIM of $\rho^{*}$ assumes the form
\begin{equation}
    \mathcal{F}_{\rho^{*}} = m_1 \mathcal{F}_{\rho_1} + m_2 \mathcal{F}_{\rho_2},
\end{equation}
where the dependency in $\theta$ is omitted for clarity.

This analysis is extended to the general case as follows. Suppose that a given estimation protocol uses a set of $S$ distributed states $\{\rho_1(\theta), \ldots, \rho_{S}(\theta)\}$, and a set of $P_i$ projective measurements $\{\varpi_{1}^{i},\ldots, \varpi_{P_i}^{i}\}$ for each state $\rho_i(\theta)$. By mapping each state-measurement pair to a state, we can represent the set of distributed states as 
\begin{equation}
    \varrho = \{\rho_{11}(\theta), \rho_{12}(\theta), \ldots, \rho_{SP_{S}}(\theta)\}, \label{eq:varrho}
\end{equation}
which contains $S^{*} = \sum_iP_i$ states. In order to simplify the analysis, we change the double index in \eqref{eq:varrho} to a single index ranging from $1$ to $S^{*}$ and consider that the set of distributed states has the form
\begin{equation}
    \varrho = \{\rho_{1}(\theta), \ldots, \rho_{S^{*}}(\theta)\}.
\end{equation}
Furthermore, let $m_{i}$ denote the number of copies of $\rho_{i}(\theta)$ that are measured for estimation. The density matrix representing the combination of copies is given by
\begin{equation}
    \rho^{*} = \bigotimes_{i = 1}^{S^{*}}\rho_i(\theta)^{\otimes m_{i}},
\end{equation}
and its QFIM has the form
\begin{equation}
    \mathcal{F}_{\rho^{*}} = \sum_{i = 1}^{S^{*}} m_i \mathcal{F}_{\rho_{i}}. \label{eq:final_qfim}
\end{equation}

Describing the structure of the QFIM in the general case provides a direct way to compare different quantum tomography protocols. Moreover, the separability of $\rho^{*}$ facilitates the analytical description of the QFIM. Its complexity reduces to that of computing the QFIM of each $\rho_i$ in $\varrho$ and performing a weighted sum of the results. We highlight that any QNT method can be understood using a similar analysis.

\subsection{Estimators}

We now present the estimators that will be combined to define the complete QNT protocols. The estimators are written with respect to probability outcomes based on the parametric representation of states. These probabilities are themselves estimated based on datasets $\mathcal{D}$ of measurement outcomes. Thus, let $S^{\rho, B}$ denote a random variable representing the measurement outcome of a single copy of state $\rho$ in the basis $B$. A dataset $\mathcal{D}_{\rho}^{B, m}= \{s_{1}^{\rho, B},\ldots , s_{m}^{\rho,B}\}$ formed by the outcomes of $B$-basis measurements performed in $m$ copies of $\rho$ is the realization of $S_{\rho}^B$ $m$ times. If the $B$ is omitted in the superscript, the symbols refer to the case of measurements performed in the diagonal basis of density operators.

\subsubsection{M-state based estimators} The probability distribution $p_M(s, \theta)$ in \eqref{eq:multiProb} was used in~\cite{de2022quantum} to provide two vector estimates for $\theta$, that are based on two different estimates for $\theta_0$. We focus on the relationship between $\hat{\theta_j}$ and $\hat{\theta}_0$ that has the form
\begin{align}
    & \hat{\theta}_j = \frac{\hat{Pr}[S_{j-1} = 1] - \hat{\theta}_0}{1 - 2\hat{\theta}_0}, \text{ for } j > 0. \label{eq:mthetajestimate}
\end{align}
Given an estimate $\hat{\theta}_0 \neq 1/2$, $\hat{\theta}_j$ for $j > 0$ can be estimated by taking
\begin{equation}
    \hat{Pr}[S_{j-1} = 1] = 1 - \frac{1}{m} \sum_{i = 1}^{m}s_{i(j-1)}^{M},\label{eq:mjestimate}
\end{equation}
where the $j - 1$ indices appear because $\rho_M$ is an $n - 1$ qubit state.

\subsubsection{IE-state based estimators} The IE state is special in the sense that the probability distributions of measurements in the GHZ, X, and Z basis are all represented by the joint distribution of independent binary random variables. Using \eqref{eq:ghzIndependentProb}, \eqref{eq:ghzIndependentProbZ}, and \eqref{eq:ghzProbX} together yields
\begin{align}
    & \hat{\theta}_j = 1 - \frac{1}{m} \sum_{i = 1}^{m}s_{ij}^{IE}, j \in \{0,\ldots, n-1\},\label{eq:fullIndEstimators}\\
    &\hat{\theta}_j = 1 - \frac{1}{m} \sum_{i = 1}^{m}s_{i0}^{IE, Z} \oplus s_{ij}^{IE, Z}, j \in \{1,\ldots, n-1\},\label{eq:riEstimators} \\
    & \hat{\theta}_0 = 1 - \frac{1}{m} \sum_{i = 1}^{m}\beta(s_{ij}^{IE, X}), \label{eq:rootEstimators}
\end{align}
where $\oplus$ denote the addition module-two operator.

\subsubsection{RI estimators} The RI estimators follow \eqref{eq:riEstimators}, since the probability distribution in \eqref{eq:rootIndependentProb} is also a joint probability of independent variables that does not depend on $\theta_0$, and $\rho_{RI}$ is diagonal in the $Z$ basis.

\subsubsection{BF estimators} From \eqref{eq:ghzProbX}, estimators for $\hat{\theta}_0$ using $X$-basis measurements for $\rho_{BF}$ have the form in \eqref{eq:rootEstimators}. When $Z$ basis measurements are performed, the probability distribution of outcomes reduces to $p_M$ with a normalization coefficient of $1 / 2$. Thus, using an estimate for $\hat{\theta}_0$, an estimator for $\hat{\theta}_j$ can be obtained from \eqref{eq:mthetajestimate}
by substituting $\hat{Pr}[S_j]$ with
\begin{align}
    & 1 - \frac{1}{m} \sum_{i = 1}^{m}s_{i0}^{BF, Z} \oplus s_{ij}^{BF, Z}, \text{ for } j \in \{1,\ldots, n-1\}\label{eq:bfEstimatorsZ}.
\end{align}
 For GHZ measurements, the estimators still follow \eqref{eq:mjestimate} and \eqref{eq:rootEstimators}, although the probabilities are computed from $\mathcal{D}_{BF}^{m} = \{s_{i}^{BF}\}$.

\subsection{Tomography protocols} \label{sec:tomo_protocols}
We combine the estimators to obtain multiple protocols that completely characterize bit-flip stars.

\subsubsection{One-step protocols} We discuss two one-step protocols. The first protocol was introduced in~\cite{de2022quantum}, and uses the IE distribution circuit with GHZ measurements yielding estimators following \eqref{eq:fullIndEstimators}. In particular, the IE state is distributed $m$ times and each GHZ measurement yields an observation containing information about all parameters. The second protocol uses the BF distribution algorithm $m$ times with GHZ measurements, obtaining $\hat{\theta}_0$ from \eqref{eq:rootEstimators}, and using \eqref{eq:mjestimate} for each $j \in \{1,\ldots, n - 1\}$ to obtain $\hat{\theta}_j$. Note that equations of the form \eqref{eq:mjestimate} have a singularity when $\hat{\theta}_0 = 1 / 2$ and estimators based on such equations are not well-defined when $\theta_0 = 1 / 2$.

\subsubsection{Two-step protocols}
Both IE and BF states lead to two-step protocols using measurements in the $Z$ and $X$ basis. Thus, in each case, the end-nodes distribute $m$ copies of the state, and $m / 2$ $Z$- and $X$-basis measurements are performed. In both cases, $\hat{\theta}_0$ is obtained from \eqref{eq:rootEstimators}, while $\hat{\theta}_j$ for $j \in \{1, \ldots, n - 1\}$ is obtained from \eqref{eq:riEstimators} and \eqref{eq:mjestimate} for the IE and BF states, respectively.

We combine the RI and M distribution circuits into the following two-step protocol. First, node $v_0$ is used as the root of the RI circuit to distribute $m / 2$ copies of $\rho_{RI}$, which are measured in the $Z$ basis. The copies provide estimators for $\hat{\theta}_j$ for $j \in \{1, \ldots, n - 1\}$ of the form in \eqref{eq:riEstimators}. Secondly, the M circuit is used with $v_0$ as the root, and \eqref{eq:mjestimate} is used once for each $\hat{\theta}_j$ to obtain $n - 1$ initial estimates $\hat{\theta}^{1}_{0},\ldots, \hat{\theta}^{n - 1}_0$ for $\theta_0$. The final estimate returned for $\theta_0$ assumes the form $\hat{\theta}_0 = \sum_{i = 1}^{n - 1}\hat{\theta}^{i}_0 / (n - 1).$

\subsubsection{$n$-step protocol} The RI circuit leads to the following $n$-step protocol. Let $m$ be the number of states distributed. Each end-node $v_j$ in the star is used as the root for the RI circuit $m / n$ times. When $v_j$ is the root, the state distributed does not depend on $\theta_j$. Hence, $(n - 1) m / n$ states among the total $m$ distributed depend on $\theta_j$. The estimator in \eqref{eq:riEstimators} can be combined for each $v_k \neq v_j$ to obtain an estimator of the form
\begin{align}
    \hat{\theta}_j = 1 - \frac{n}{m(n - 1)} \sum_{s \in \mathcal{D}_j} s_j,
\end{align}
where $\mathcal{D}_j$ denotes the combined dataset of the $(n - 1) m / n$ RI circuits performed with all end-nodes $v_k \neq v_j$ as root, and $s_j$ denotes the measurement bit obtained in $v_j$ once a sample is locally measured in the $Z$ basis.

\section{Evaluation}\label{sec:evaluation}

The six protocols presented in Section~\ref{sec:tomgoraphy} depict the diverse space of solutions for the tomography of quantum bit-flip networks. In this section, we numerically evaluate and compare the performance of all six protocols discussed.
\begin{figure*}
    \begin{centering}
        \hfill
        \subfloat[QFIM's inverse trace per parameter value.\label{subfig:cov_tr}]{\includegraphics[scale=0.55]{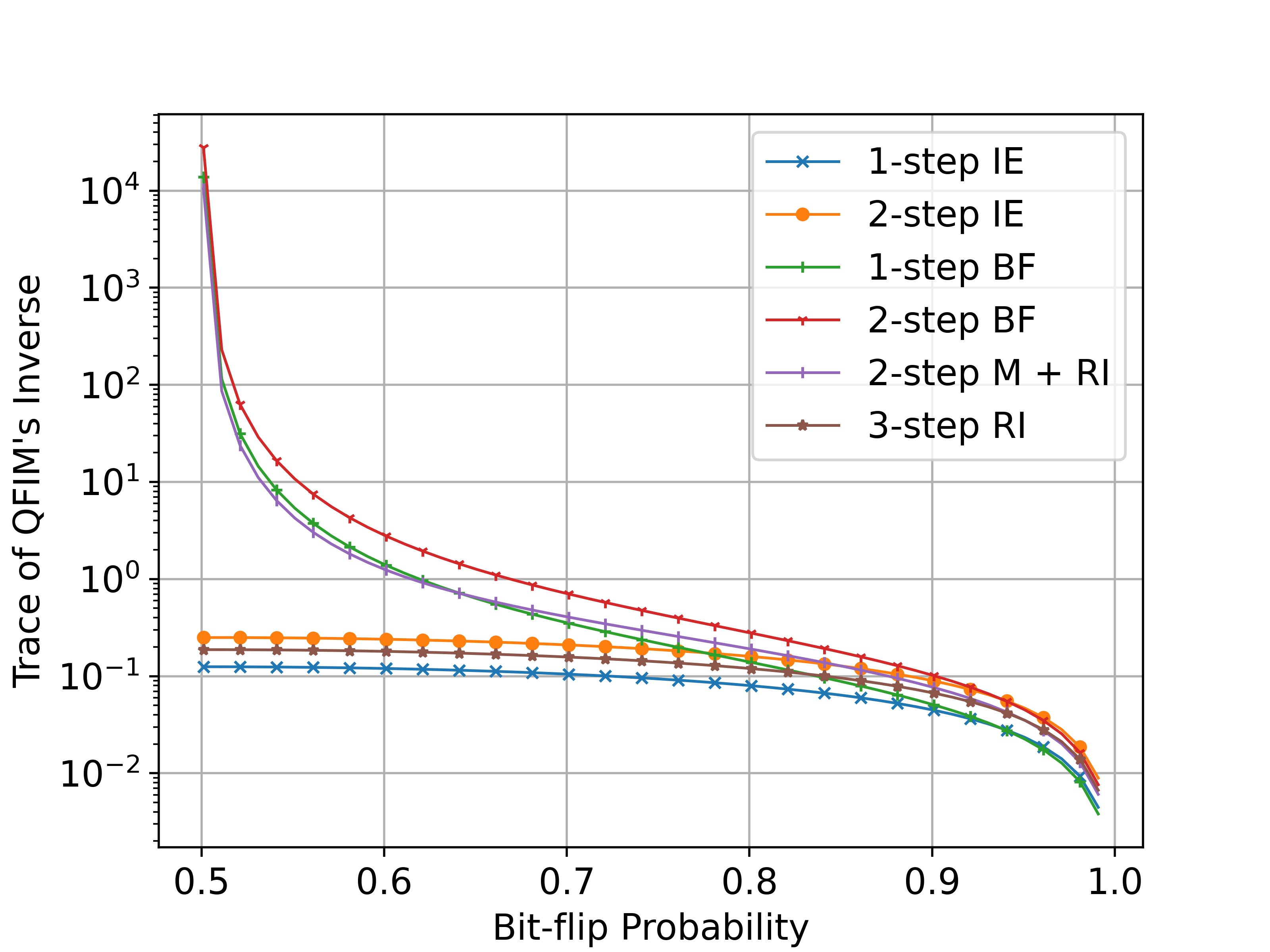}} \hfill
        \subfloat[Estimation error per number of samples.\label{subfig:error}]{\includegraphics[scale=0.55]{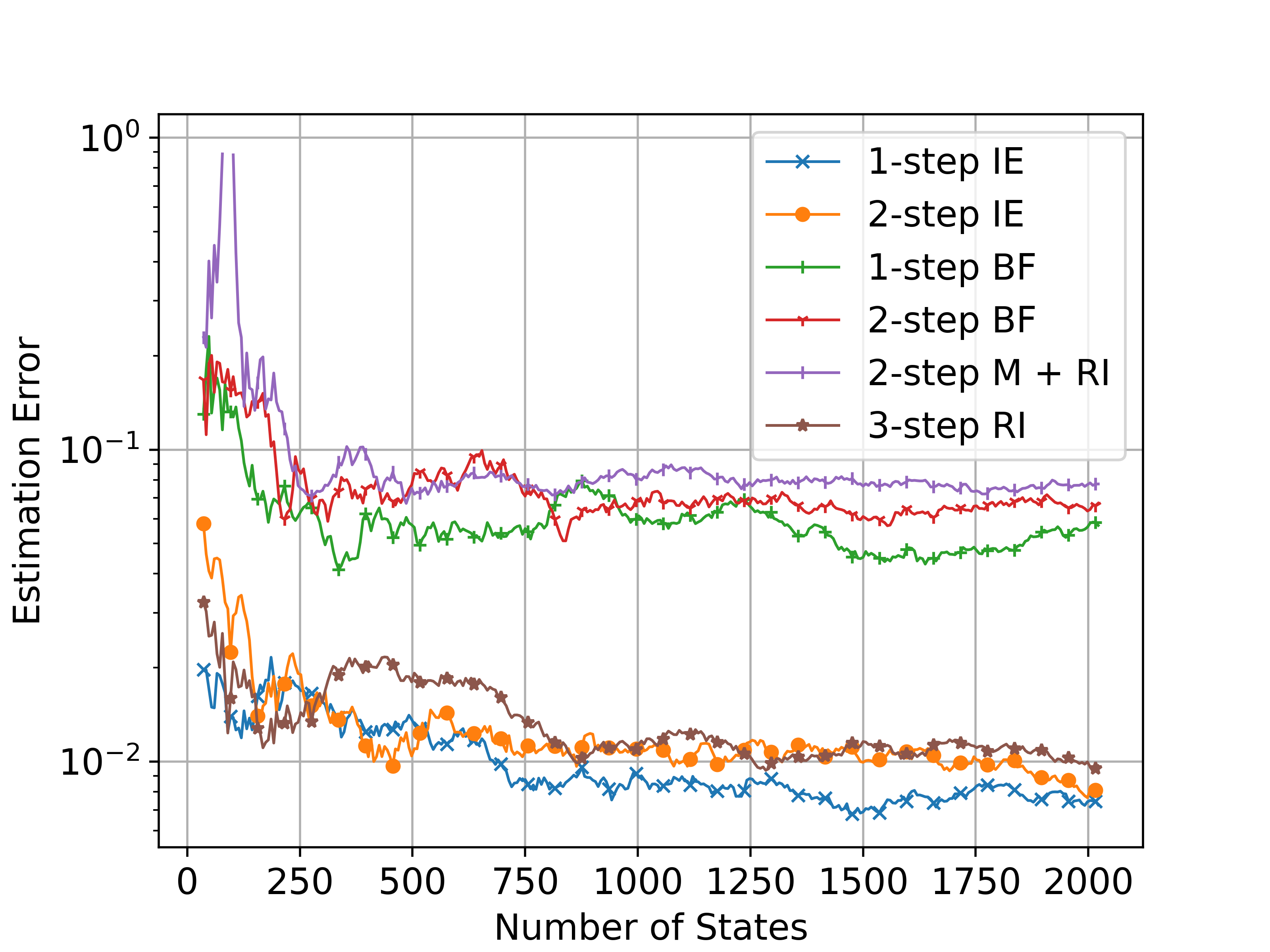}} \hfill
    \end{centering}
    \caption{Numerical evaluation of estimation performance for the QNT protocols presented in Section~\ref{sec:tomo_protocols} when applied to the characterization of a four-node star with channels of equal bit-flip probability $\theta^{*}$. \textbf{(\subref{subfig:cov_tr})} Trace of $\mathcal{F}^{-1}$ per $\theta^{*}$. Since the QFIMs are symmetric with respect to 0.5, we only show the second half of the parameter region. Each protocol distributes a total of $m = 6$ states to estimate $\theta$, which is the least common multiple of the minimum number of states required for the protocols to estimate $\hat{\theta}$.
    \textbf{(\subref{subfig:error})} Estimation error $\norm*{\hat{\theta} - \theta^{*}}$ per number of states used for estimation for a bit-flip star with $\theta^{*} = 0.58$, averaged over 5 trials. The $x$-axis varies in increments of $6$, ranging from $m = 36$ to $m = 2024$.}\label{fig:qfim}
\end{figure*}

We start with a numerical analysis of the QFIM inverse for each tomography protocol. The QCRB, \eqref{eq:cramer_rao}, implies that the trace of the QFIM's inverse is a lower bound on the sum of the variances of estimators. In particular, $\Tr[\mathcal{F}^{-1}]$ lower bounds the sum of the variances of entries $\hat{\theta}_j$ of any estimator $\hat{\theta}$ obtained from measurements of $\rho$. The states distributed for tomography have QFIMs following \eqref{eq:qfim_def} and, as discussed in Section~\ref{sec:tomgoraphy}, the combined copies of states and measurements are captured by QFIMs following \eqref{eq:final_qfim}. Let $\theta^{*} \in [0, 1]$ denote a fixed probability value. For each QNT protocol $\mathcal{P}$, we compute $\Tr[ \mathcal{F}^{{*-1}}]$ when $m$ states are distributed to obtain one estimate $\hat{\theta}$ of the entire parameter vector $\theta$, for a four-node star with uniform bit flip probability, i.e., $\theta_j = \theta^{*}$ for $j \in \{0, 1, 2\}$. An $n$-step protocol requires $n$ states to be distributed to obtain a single estimate $\hat{\theta}$. Therefore, we use $m = 6$ for every protocol, since six is the least common multiple of the number of states required by the protocols in a four-node star.

Our results are reported in Figure~\ref{subfig:cov_tr}. Interestingly, the curves highlight that the relative behavior of the inverse trace changes based on $\theta^{*}$. In particular, BF- and M-based protocols exhibit lower QCRBs when $\theta^{*}$ is far from $1 / 2$, and large bounds when close, while RI- and IE-based protocols show a smoother relationship with $\theta^{*}$. Furthermore, the one-step BF protocol yields the lowest QCRB when $\theta^{*}$ is either close to zero or one, while the one-step IE protocol has the lowest bound for most of the parameter regime. The curves also highlight the advantages provided by entanglement in estimation vary according to the parameter values, which has been previoulsy reported in~\cite{fujiwara2001quantum} for the point-to-point estimation of depolarizing channels.

To further our analysis, we simulate the QNT protocols using Netsquid~\cite{coopmans2021netsquid}. We study four-node networks with $\theta^{*} = 0.58$, and compute the norm $\norm*{\hat{\theta} - \theta^{*}}$ to analyze the convergence behavior of the estimators with the number of distributed states used for estimation. Figure.~\ref{subfig:error} shows results for the case when the number of states used by each protocol is varied from 36 to 2024, averaged over five trials. As expected, two distinct groupings appear, with the BF- and M- based protocols reporting a significantly higher variance than the rest of the protocols. This grouping agrees with our theoretical expectations given the QCRBs shown in Figure~\ref{subfig:cov_tr} for $\theta^{*} = 0.58$. The combined evaluation of the QCRBs and the convergence behavior of estimators is a first step towards the rigorous benchmarking of QNT protocols. Our findings provide evidence that the performance of QNT methods can depend on the values of parameters to be estimated, offering initial insights to the design of optimal QNT protocols. In particular, it paves the way for adaptive QNT methods that dynamically modify estimation strategy based on current parameter estimates in order to exploit the distinct efficiencies of protocols with parameter values.


\section{Conclusion}\label{sec:conclusion}
The results reported in this article further the end-to-end characterization of bit-flip quantum stars. We reviewed the methods proposed in~\cite{de2022quantum} and provided novel QNT protocols that utilize multiple state distribution protocols and measurements. The proposed protocols uniquely characterize bit-flip probabilities in quantum star networks by exploiting both local and global network measurements, achieving varying estimation efficiency. Moreover, the QFIM analysis presented in this article is general and provides new insights in the design of QNT protocols. The numerical evaluation of the trace of the QFIM's inverse shed light on entanglement advantages for QNT. In particular, our findings show that the proposed QNT protocols which do not rely on global measurements exhibit comparable performance to the ones that use pre-shared entanglement to perform measurements at the end-nodes. Thus, determining the conditions under which entanglement yields optimal QNT methods and provides significant quantum advantage is a fundamental research question that we identify as future work. Furthermore, the results presented in this article are stepping stones toward the goal of devising QNT protocols for the characterization of quantum star networks formed by arbitrary Pauli channels. Uniquely determining probabilities for this general case is considerably harder than the bit-flip scenario considered in this work. Nonetheless, the state distribution schemes proposed serve as inspiration to the inquiry of efficient parameter encoding schemes for the characterization of generic Pauli noise.

We identify the analysis of QNT protocols under the assumption of imperfect network hardware as a key direction for future work. The protocols developed in this article assume that nodes have access to perfect quantum operations and memories, disregarding the effects of operation noise in the end-to-end estimation of link parameters. Understanding the limits of end-to-end network characterization in the face of noise is fundamental for the development of useful QNT protocols, as it can help guide the development of quantum network management tools to inform protocol designers and network managers.


{\em Acknowledgments}---This research was supported in part by the NSF grant CNS-1955744, NSF- ERC Center for Quantum Networks grant EEC-1941583, and MURI ARO Grant W911NF2110325.

\newpage

\newpage
\bibliographystyle{unsrt}
\bibliography{references}


\end{document}